\documentclass[12pt,aps,prb,preprint]{revtex4}
\usepackage{color}

\begin{document}
\def\beq{\begin{eqnarray}}
\def\eeq{\end{eqnarray}}
\newcommand{\nn}{\nonumber}
\newcommand{\reals}{\mbox{${\rm I\!R }$}}
\newcommand{\nats}{\mbox{${\rm I\!N }$}}
\newcommand{\intgs}{\mbox{${\rm Z\!\!Z }$}}
\newcommand{\ep}{\epsilon}
\newcommand{\cao}{{\cal O}}
\newcommand{\intl}{\int\limits}
\newcommand{\G}{\Gamma}
\newcommand{\bcs}{boundary conditions }
\newcommand{\pa}{\partial}
\newcommand{\com}{\left(\frac{k^2 r^2}{\nu^2} - 1\right)}
\noindent

\title{Computation of determinants using contour integrals}

\author{Klaus Kirsten}\email{klaus_kirsten@baylor.edu}
\affiliation{Department of Mathematics, Baylor University, Waco,
TX 76798}
\author{Paul Loya}\email{paul@math.binghamton.edu}
\affiliation{Department of Mathematics, Binghamton University,
Vestal Parkway East, Binghamton, NY 13902}


\begin{abstract}
It is shown how the pre-exponential factor of the Feynman
propagator for a large class of potentials can be computed using
contour integrals. This is of direct relevance in the context of
tunnelling processes in quantum theories. The prerequisites for
this analysis are accessible to advanced undergraduate students
and involve only introductory courses in ordinary differential
equations and complex variables.
\end{abstract}

\maketitle

\section{Introduction}
The spectrum of certain differential operators encodes fundamental
properties of different physical systems. Various functions of the
spectrum, the so-called spectral functions, are needed to decode
these properties. One of the most prominent spectral functions is
the zeta function, which relates for example to partition sums,
the heat-kernel and the functional determinant; see, e.g., Ref.
16\nocite{kirs02b}. Zeta functions are often associated with
suitable sequences of real numbers $\{\lambda_k\}_{k\in\nats}$,
which, for many applications, are eigenvalues of Laplace-type
operators. In generalization to the zeta function of Riemann, \beq
\zeta_R (s) = \sum_{k=1}^\infty k^{-s},\label{1}\eeq one defines
\beq \zeta (s) = \sum_{k=1}^\infty \lambda_k^{-s}, \label{2}\eeq
where $s$ is a complex parameter whose real part is assumed to be
sufficiently large such as to make the series convergent.

To indicate how zeta functions relate to other spectral functions,
let us use the functional determinant as an example because it is
going to be the focus of the article. For the purpose of relating
zeta functions and determinants assume for the moment that we talk
about a sequence of {\it finitely} many numbers
$\{\lambda_k\}_{k=1}^n$. Considering them as eigenvalues of a
matrix $L$, we have \beq \det L = \prod_{k=1}^n \lambda_k ,
\nn\eeq which implies $$\left. \ln \det L = \sum_{k=1}^n \ln
\lambda_k = - \frac d {ds} \right|_{s=0} \sum_{k=1}^n
\lambda_k^{-s}.$$ In the notation of Eq. (\ref{2}) this shows \beq
\ln \det L = - \zeta ' (0) \quad \mbox{or} \quad \det L =
e^{-\zeta ' (0)} . \label{3} \eeq When the finite dimensional
matrix is replaced by a differential operator $L$ having
infinitely many eigenvalues, in general $\prod_{k=1}^\infty
\lambda_k$ will not be defined. However, as it turns out for many
situations of relevance, definition (\ref{3}) makes perfect sense
and has found important applications in mathematics and physics;
for the first appearance of definition (\ref{3}) see Refs. 10, 13,
and 24\nocite{dowk76-13-3224,hawk77-55-133,ray71-7-145}.

In recent years a contour integral method has been developed for
the analysis of zeta functions, \cite{bord96-37-895,kirs02b}
which, although applicable in any dimension and to a variety of
spectral functions, shows its full elegance and simplicity when
applied in one dimension, and when applied to functional
determinants. One of the main reasons for the relevance of
determinants is the fact that the evaluation of the Feynman
propagator involves this quantity. The probably most important
field of application of functional determinants deals with
tunnelling processes in quantum mechanics, quantum field theory
and quantum statistics. \cite{klei06b,schu81b} Because of its
relevance a considerable number of articles in American Journal of
Physics have been devoted to this topic; see, e.g., Refs. 1, 5, 7,
14, and
15\nocite{albu98-66-524,baro01-69-232,brow94-62-806,hols97-65-414,hols98-66-583},
and so we decided to also concentrate on this topic. Our aim is to
show how and why a contour integral method is extremely well
adapted for the evaluation of {\it in particular} functional
determinants. By demonstrating an additional way by which results
may be obtained we enlarge the arsenal of techniques by a
component already proven useful in recent research, see, e.g.,
Refs. 17 and 19\nocite{kirs04-37-4649,kirs06-321-1814}. The
probably most attractive feature of our approach is that all
prerequisites are known to advanced undergraduate students of
physics and mathematics. Namely, we only assume some working
knowledge with Cauchy's residue theorem \cite{conw78b} and some
elementary facts about ordinary differential equations.
\cite{nagl04b}

The outline of this article is as follows. We explain the basic
ideas of our approach by looking at the zeta function of Riemann,
and by evaluating $\zeta_R ' (0)$. This is identical to the
evaluation of the functional determinant of a free particle in an
interval with Dirichlet boundary conditions at the endpoints. We
then will consider the case of particles in a harmonic oscillator
potential previously considered in Refs. 1, 7, and
15\nocite{albu98-66-524,baro01-69-232,hols98-66-583}. Results will
be trivially rederived. Finally, we show how particles moving in
any potential (satisfying reasonable conditions) and obeying quite
general boundary conditions can be analyzed. The Conclusions
highlight the most important points of our contribution.
\section{Functional determinant of a free particle in an interval}
A free particle in an interval is described by the operator $\frac
{d^2}{dt^2}$ together with some boundary condition. In the context
of the Feynman propagator Dirichlet boundary conditions are quite
common \cite{hols98-66-583} and this is what we first concentrate
on. It will be convenient to make a rotation in the complex
$t$-plane and to define $t=-i\tau$. The resulting operator
\[
P= - \frac{d^2}{d\tau^2}
\]
in terms of $\tau$ has positive eigenvalues.  This is the relevant
setting in the context of quantum tunnelling,
\cite{LanJ67,LanJ69,ColS77,CaCCoS77,klei06b,schu81b} cf.\ the last
part of Section \ref{sec-genpotential}.

So in order to evaluate the functional determinant associated with
this situation we consider the eigenvalue problem $$- \frac{d^2}
{d\tau^2} \phi _n (\tau ) = \lambda_n \phi _n (\tau ), \quad \quad
\phi_n (0) = \phi _n (L) = 0.$$ The eigenfunctions have the form
$$\phi_n (\tau ) = a \sin (\sqrt{\lambda_n }\, \tau ) + b \cos
(\sqrt{\lambda_n}\,  \tau ).$$ The appearance of the cosine is
excluded by the boundary value $\phi_n (0) = 0$. The eigenvalues
are then found from the equation \beq\sin (\sqrt { \lambda_n } L )
=0. \label{4}\eeq This condition is simple enough to be solved for
analytically and one determines $$\phi_n (\tau ) = a \sin
(\sqrt{\lambda_n} \tau ) , \quad \quad \lambda_n = \left(
\frac{n\pi } L \right)^2, \quad n\in\nats,$$ with some
normalization constant $a$.

Although in this particular case it is of course convenient to
have an explicit expression for the eigenvalues, let us pretend
the best we are able to obtain is an equation like (\ref{4}),
namely eigenvalues are determined as the zeroes of some function
$F(\lambda )$. As we will see, actually this is as convenient as
having explicit eigenvalues,  but of much larger applicability.

For the given setting the natural choice $F(\lambda ) = \sin
(\sqrt{\lambda} L )$ has to be modified as $\lambda =0$ satisfies
$F(0) = 0$. In order to avoid $F(\lambda )$ having more zeroes
than there are actual eigenvalues we therefore define \beq
F(\lambda ) = \frac{\sin (\sqrt \lambda L)} {\sqrt \lambda} .
\label{5}\eeq Note that $F(\lambda)$ is an entire function of
$\lambda$. The next step in the contour integral formalism is to
rewrite the zeta function using Cauchy's integral formula. Given
that $F(\lambda )=0$ defines the eigenvalues $\lambda_n$, then
$$\frac d {d\lambda} \ln F(\lambda ) = \frac{F' (\lambda )}{F
(\lambda )}$$ has poles exactly at those eigenvalues. Furthermore,
expanding about $\lambda = \lambda _n$ we see for $F' (\lambda _n)
\neq 0$ that $$\frac{F' (\lambda )} { F (\lambda )} = \frac{ F'
(\lambda - \lambda_n + \lambda_n )}{F (\lambda - \lambda_n +
\lambda _n )} = \frac{ F ' (\lambda _n) + (\lambda - \lambda_n) F
'' (\lambda_n ) + ...} { (\lambda - \lambda_n) F ' (\lambda ) + (
\lambda  - \lambda_n ) ^2 F '' (\lambda_n) + ...} = \frac 1
{\lambda - \lambda_n } + ...$$ and the residue at all eigenvalues
is $1$. (A variation of this argument shows that if $m_n$ is the
multiplicity of $\lambda_n$, the residue of $F '(\lambda)/F
(\lambda )$ at $\lambda_n$ is $m_n$.) This shows, noticing the
appropriate behavior of $F(\lambda)$ at infinity, that for $\Re s
> \frac 1 2$, \beq \zeta _P (s) = \frac 1 {2\pi i}
\int\limits_\gamma d\lambda \,\, \lambda^{-s} \frac d {d\lambda}
\ln F(\lambda) , \label{6} \eeq where the contour $\gamma$ is
shown in Figure 1.

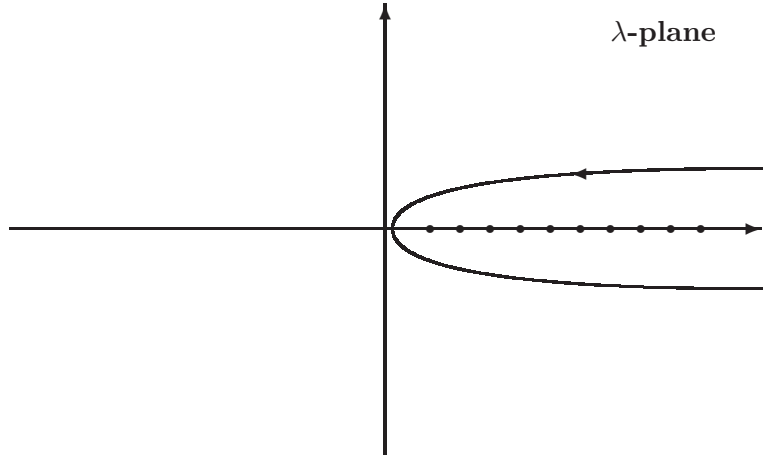
\begin{figure}[ht]
\setlength{\unitlength}{1cm}

\begin{center}
\begin{picture}(10,6.5)
\thicklines

\put(0,3){\vector(1,0){10}} \put(5.0,0){\vector(0,1){6}}
\multiput(5.6,3)(.4,0){10}{\circle*{.1}}

\put(8.0,5.5){{\bf $\lambda$-plane}}

\qbezier(5.1,3)(5.1,3.8)(10,3.8) \qbezier(5.1,3)(5.1,2.2)(10,2.2)

\put(7.5,3.735){\vector(-1,0){0.036}}

\end{picture}
\caption{\label{fig1}Contour $\gamma$}
\end{center}
\end{figure}

As is typical for complex analysis, the next step in the
evaluation of a line integral is a suitable deformation of the
contour. Roughly speaking, deformations are allowed as long as one
does not cross over poles or branch cuts of the integrand. For the
integrand in (\ref{6}), the poles are on the positive real axis,
and there is a branch cut of $\lambda^{-s}$ which we define to be
on the negative real axis, as is customary. So as long as the
bahavior at infinity is appropriate, we are allowed to deform the
contour to the one given in Figure 2.

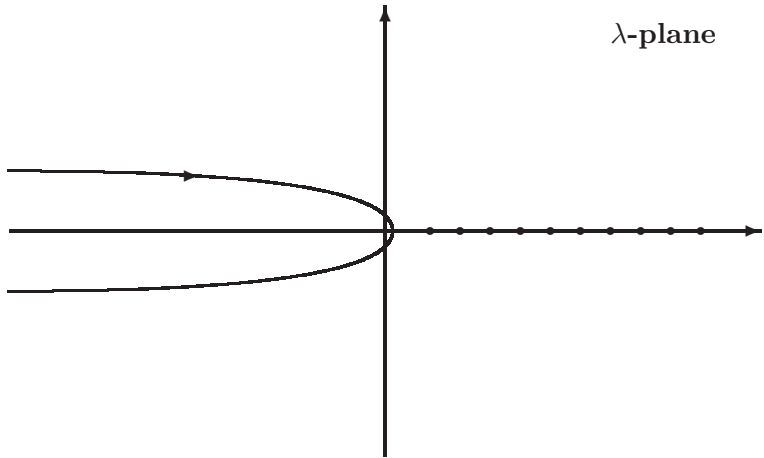
\begin{figure}[ht]
\setlength{\unitlength}{1cm}

\begin{center}
\begin{picture}(10,6.5)
\thicklines

\put(0,3){\vector(1,0){10}} \put(5.0,0){\vector(0,1){6}}
\multiput(5.6,3)(.4,0){10}{\circle*{.1}}

\put(8.0,5.5){{\bf $\lambda$-plane}}

\qbezier(5.1,3)(5.1,2.2)(0,2.2) \qbezier(5.1,3)(5.1,3.8)(0,3.8)

\put(2.5,3.735){\vector(1,0){0.036}}

\end{picture}
\caption{\label{fig2}Contour $\gamma$ after deformation}
\end{center}
\end{figure}

\begin{figure}[ht]
\setlength{\unitlength}{1cm}

\begin{center}
\begin{picture}(10,6.5)
\thicklines

\put(0,3){\vector(1,0){10}} \put(5.0,0){\vector(0,1){6}}
\multiput(5.6,3)(.4,0){10}{\circle*{.1}}

\put(8.0,5.5){{\bf $\lambda$-plane}}

\qbezier(5,3.01)(4,3.01)(0,3.01) \qbezier(5,2.99)(4,2.99)(0,2.99)
\qbezier(5,2.97)(4,2.97)(0,2.97) \qbezier(5,3.03)(4,3.03)(0,3.03)

\put(2.2,3.735){\vector(1,0){1.536}}
\put(3.7,2.265){\vector(-1,0){1.536}}

\put(2.2,3.9){$\lambda = e^{i \pi} x$}

\put(2.1,1.85){$\lambda = e^{-i \pi} x$}

\end{picture}
\caption{\label{fig3}Contour $\gamma$}
\end{center}
\end{figure}
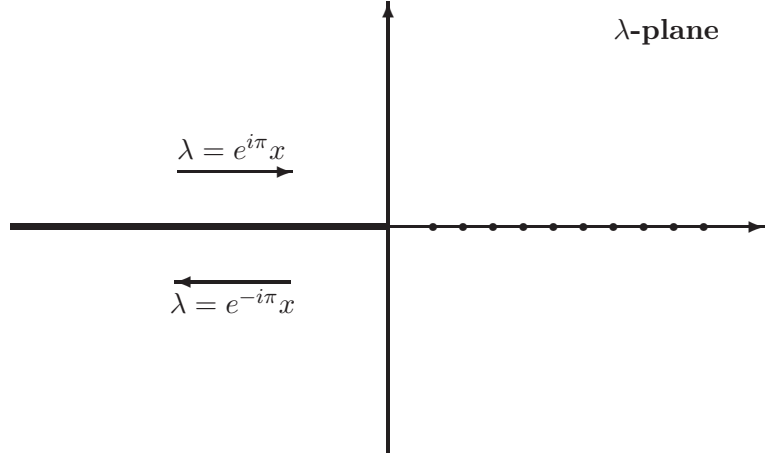

In order to better see the $|\lambda | \to \infty$ behavior of $F
(\lambda )$, let us rewrite the sine in terms of exponentials. We
then have $$F(\lambda ) = \frac 1 {2i\sqrt \lambda } \left(
e^{i\sqrt \lambda L} - e^{-i \sqrt \lambda L}\right), $$ and for
$\Re s > \frac 1 2$ all deformations are indeed allowed. We next
want to shrink the contour to the negative real axis as shown in
Figure \ref{fig3}.

As $\lambda$ approaches the negative real axis from above,
$\lambda ^{-s}$ picks up a phase $(e^{i\pi})^{-s} = e^{-i\pi s}$,
whereas the limit from below produces $(e^{-i\pi })^{-s} = e^{i\pi
s}$. Given the opposite direction of the contour above and below
the negative real axis, contributions add up to produce a $\sin
(\pi s)$. Taking care of the same kind of argumentation in
$F(\lambda )$ one obtains \beq \zeta _P (s) = \frac{ \sin \pi s}
\pi \int\limits_0^\infty dx \,\, x^{-s} \frac d {dx} \ln \left(
\frac{ e^{\sqrt x L}} {2 \sqrt x} \left[ 1 - e^{-2 \sqrt x L}
\right]\right) . \label{7} \eeq Notice, that by shrinking the
contour to the negative real axis a new condition for the integral
to be well defined, namely $\Re s < 1$, has become necessary due
to the behavior about $x=0$.

Let us rest for a moment to stress the nice features of equation
(\ref{7}) for the evaluation of determinants. If the integral were
finite at $s=0$, an evaluation  of the determinant would be
trivial. From

\beq \zeta _P '(0) &=& \left(\frac{d}{ds} \Big|_{s = 0} \frac{
\sin \pi s}{\pi} \right) \cdot \left( \int\limits_0^\infty dx \,\,
x^{-s} \frac d {dx} \ln \left( \frac{ e^{\sqrt x L}} {2 \sqrt x}
\left[ 1 - e^{-2 \sqrt x L} \right]\right) \right)\Bigg|_{s = 0} \nn\\
& &+ \left( \frac{ \sin \pi s} \pi \right)\Big|_{s = 0} \cdot
\left( \frac{d}{ds} \Big|_{s = 0} \int\limits_0^\infty dx \,\,
x^{-s} \frac d {dx} \ln \left( \frac{ e^{ \sqrt x L}} {2 \sqrt
x} \left[ 1 - e^{-2  \sqrt x L} \right]\right) \right)\nn \\
& = & \int\limits_0^\infty dx \,\, \frac d {dx} \ln \left( \frac{
e^{ \sqrt x L}} {2 \sqrt x} \left[ 1 - e^{-2 \sqrt x L}
\right]\right)  \label{simple}\eeq it would amount to finding $\ln
(...)$ at the limits of integration; even no integration needed to
be done explicitly. Whereas this is exactly what occurs when
considering {\it ratios} of determinants, see Section 4, for
absolute determinants the situation is slightly more complicated.
The reason is that (\ref{7}) is only well defined for $\frac 1 2 <
\Re s <1$ and a little more effort is needed. Notice, the problem
is caused by the $x\to\infty$ behavior which enforces the
condition $\frac 1 2 < \Re s$. To analyze this further we
therefore split the integral according to $\int_0^1 dx +
\int_1^\infty dx$. Whereas from the above remarks it follows that
$\int_0^1 dx$ can be considered to be in final form, the
$\int_1^\infty dx$ needs further manipulation. The pieces needing
extra attention are \beq \int\limits_1^\infty dx \,\, x^{-s} \frac
d {dx} \ln e^{ \sqrt x L} &=& \frac { L} 2 \int\limits_1^\infty dx
\,\, x^{-s-\frac 1 2} =
\frac { L} {2s-1} , \nn\\
\int\limits_1^\infty dx \,\, x^{-s} \frac d {dx} \ln \left( \frac
1 {2 \sqrt x} \right) &=& - \frac 1 2 \int\limits_1^\infty dx \,\,
x^{-s-1} = - \frac 1 {2s} . \nn\eeq This shows \beq \zeta _P (s)
&=& \frac{L \sin \pi s} {(2s-1)\pi} - \frac{\sin \pi s} {2s \pi} +
\frac {\sin \pi s} \pi \int\limits_1^\infty dx \,\, x^{-s} \frac d
{dx}
\ln \left( 1 - e^{-2 \sqrt x L}\right) \nn\\
& &+\frac {\sin \pi s} \pi \int\limits_0^1 dx \,\, x^{-s} \frac d
{dx} \ln \left( \frac{ e^{ \sqrt x L} } {2\sqrt x} \left[ 1 -
e^{-2 \sqrt x L}\right]\right), \nn\eeq a form perfectly suited
for the evaluation of $\zeta_P ' (0)$. We find $$\zeta _P ' (0) =
-  L- 0 - \ln \left( 1-e^{-2 L}\right) + \ln \left( \frac {e^{ L
}} 2 \left[ 1-e^{-2 L } \right] \right) - \ln L = - \ln (2L) .$$
This, of course, agrees with the answer found from the well known
values $\zeta _R (0) = - \frac 1 2 $, $\zeta _R ' (0) = - \frac 1
2 \ln (2\pi )$: \beq & &\zeta _P (s) = \sum_{n=1} ^\infty \left(
\frac {n\pi } L \right)^{-2s} =  \left( \frac L \pi
\right) ^{2s} \zeta _R (2s) \nn\\
& &\Longrightarrow \zeta _P ' (0) = 2 \ln \left( \frac L \pi
\right) \zeta _R (0) + 2 \zeta _R ' (0) = - \ln \left( \frac L \pi
\right) - \ln (2\pi) = - \ln (2L).\label{8}\eeq
\section{Functional determinant for particles in a harmonic
oscillator potential} Let $\omega$ be the frequency of the
harmonic oscillator, then the relevant operator to be considered
is $$ P_{ho} = - \frac{d^2}{d\tau ^2} + \omega ^2 , $$ with
Dirichlet boundary conditions imposed at the endpoints $\tau = 0 $
and $\tau = L$. Eigenvalues are then determined by the implicit
equation \beq \sin (\sqrt{\lambda_n  - \omega ^2} L ) = 0 .
\label{9} \eeq Instead of looking at the determinant of $P_{ho}$
itself, let us now consider the ratio $\det (P_{ho}) /\det( P)$,
where as before $P = - \frac{d^2}{d \tau^2}$, that is we consider
the difference of the associated zeta function. Using the same
strategy as before in Section 2, we have
$$ \zeta _{P _{ho}} (s) - \zeta _P (s) = \frac 1 {2\pi i}
\int\limits_\gamma d\lambda \,\, \lambda^{-s} \frac d {d\lambda}
\ln \left( \frac{ \sin (\sqrt{\lambda - \omega ^2}L)} {\sin (\sqrt
\lambda L)} \frac {\sqrt \lambda}{\sqrt {\lambda - \omega^2}}
\right),$$ the contour $\gamma$ still given by Figure 1. Deforming
as before, we obtain \beq \zeta _{P _{ho}} (s) - \zeta _P (s) =
\frac{\sin \pi s} \pi \int\limits_0^\infty dx \,\, x^{-s} \frac d
{dx} \ln \left( \frac{\sinh (\sqrt{x+\omega^2} L)}{\sinh (\sqrt x
L )} \frac {\sqrt x}{\sqrt{x+\omega^2}}\right), \label{10} \eeq
where $\sin (iy) = i \sinh y$ has been used. The technically
simplifying consequence of considering ratios gets now apparent:
as $x$ tends to infinity, the behavior of the integrand has
improved. In detail we have as $x\to\infty$ \beq \frac{\sinh
(\sqrt {x+\omega^2} L)} {\sinh (\sqrt x L) } \frac{\sqrt x}{\sqrt
{x+\omega^2}} = e^{L (\sqrt {x+\omega^2} - \sqrt x )} \frac{ \sqrt
x } { \sqrt {x+\omega^2}} \frac {1-e^{-2 L
\sqrt{x+\omega^2}}}{1-e^{-2 L \sqrt x}} = 1 + \frac 1 2
\frac{\omega^2 L } { \sqrt x} + ...\nn\eeq and the integrand
behaves like $x^{-s-3/2}$. Noting that the $x\to 0$ behavior up to
a proportionality constant is as before, we see that (\ref{10}) is
well defined for $-\frac 1 2 < \Re s < 1$, in particular, it is
well defined at $s=0$. Thus, trivially, following along the lines
leading to (\ref{simple}), $$ \zeta ' _{P_{ho}} (0) - \zeta ' _P
(0) = - \ln \left( \frac{ \sinh \omega L }{\omega L } \right) , $$
or, switching back to real time, replacing $L= i (t_f - t_i)$,
\beq \ln \frac{\det P_{ho}}{\det P} = \ln \frac{\sinh (i\omega
(t_f - t_i))} {i\omega (t_f - t_i)} = \ln \frac{ \sin (\omega (t_f
- t_i ))}{\omega (t_f - t_i )}, \label{11}\eeq the well known
answer; see, e.g., Refs. 11 and 15\nocite{feyn65b,hols98-66-583}.

Other boundary conditions can be dealt with basically with no
extra effort. For example let us consider quasi-periodic boundary
conditions as they have been analyzed for anyon-like oscillators.
\cite{bosc95-205-255,bosc95-28-7} In this case the boundary
condition reads $$\phi_n (L) = e^{i\theta } \phi_n (0), \quad
\quad \phi_n ' (L) = e^{i\theta } \phi_n ' (0) , $$ with $\theta$
some real parameter; $\theta =0$ corresponds to periodic boundary
conditions, whereas $\theta = \pi$ gives antiperiodic boundary
conditions typical for fermions. The general form of
eigenfunctions is $$ \phi _n (\tau ) = a \sin \left(
\sqrt{\lambda_n - \omega^2 } \tau \right) + b \cos \left( \sqrt{
\lambda_n - \omega ^2 } \tau \right). $$ The boundary condition
produces the equations, use $\mu _n = \sqrt{\lambda_n-\omega^2}$,
$$ a \sin (\mu _n L) + b \cos (\mu _n L) = e^{i\theta } b, \quad \quad
- \mu _n b \sin (\mu _n L) + \mu_n a \cos (\mu_n L) = e^{i\theta }
\mu_n a .$$
Under the assumptions that $\mu_n \ne 0$, which excludes periodic
boundary conditions, this system represents the matrix equation
$$
\left(\begin{array}{cc} \sin(\mu_n L) & \cos(\mu_n L) - e^{i\theta} \\
\cos(\mu_n L) - e^{i\theta}  & - \sin(\mu_n L)
\end{array}\right) \left(\begin{array}{c} a \\ b
\end{array}\right) = 0.
$$
This has a nontrivial solution if and only if the determinant of
the matrix is zero, which after some simple manipulations gives
the condition for eigenvalues as
$$ \cos (\mu _n L) - \cos \theta =0.$$ Following the steps of the
previous calculation, denoting the operator with quasi-periodic
boundary conditions as $P_{ho}^{qp}$ and $P^{qp}$, the answer can
essentially be simply read off, $$ {\zeta_{P_{ho}} ^{qp}}' (0) -
{\zeta _P ^{qp}} ' (0) = - \ln \left( \frac{ \cosh (\omega L) -
\cos \theta }{1-\cos \theta } \right) , $$ and agrees with Ref.
3\nocite{bosc95-205-255}. So in real time, $$\ln \frac{\det
P_{ho}^{qp}}{\det P^{qp}} = \ln \frac{\cos (\omega (t_f - t_i)) -
\cos \theta }{1-\cos \theta } .$$ For periodic boundary conditions
an eigenfunction with zero eigenvalue occurs, namely the constant,
and we comment on this situation  in the conclusions.
\section{Functional determinants of particles in general
potentials} \label{sec-genpotential} As the previous section made
clear, the answer was obtained without ever worrying what the
actual eigenvalues of the operator in question might be. The only
information that entered was the implicit eigenvalue equation
(\ref{9}). Is there any way an equation like (\ref{9}) can be
obtained for general potentials, such that the evaluation of
determinants is similarly trivial as the previous one? The answer
is yes and elementary knowledge of ordinary differential equations
is all that is needed. \cite{nagl04b}

So let us say we were interested in the ratio of determinants of
operators of the type $$ P_j = - \frac {d^2}{d\tau ^2} + R_j (\tau
), \quad \quad j=1,2,$$ where for convenience again Dirichlet
conditions are considered. In the previous sections $R_2 (\tau )
=0$ was chosen, but no additional complication arises for this
more general case. Such ratios arise, for example, in the
evaluation of decay probabilities in the theory of quantum
tunnelling. \cite{LanJ67,LanJ69,ColS77,CaCCoS77} Recall that if a
quantum particle moves in a potential $V(x)$ for which classically
a particle is at rest at $x = 0$, and if $\overline{x}$ denotes
the, say only, stationary point of the Euclidean action, then to
leading order in $\hbar$ the decay probability per unit time of
the unstable state is a multiple of, see Eq. (2.25) of Ref.
9\nocite{CaCCoS77},
\[
\left| \frac{\displaystyle \det \left( - \frac {d^2}{d\tau ^2} +
V''(\overline{x}) \right)}{\displaystyle \det \left( - \frac
{d^2}{d\tau ^2} + V''(0) \right)} \right|^{-1/2}.
\]
Our contour integration method can easily handle such ratios.
Indeed, as suggested by the previous examples, in order to
evaluate $\det P_1/\det P_2$ in the general case, the contour
integral to be written down should involve solutions to the
equation $$ P_j \phi _{j,\lambda} (\tau ) = \lambda
\phi_{j,\lambda} (\tau ), $$ where $\lambda$, for now, is an
arbitrary complex parameter. As is well known, for continuous
potentials $R_j (\tau )$ there will be two linearly independent
solutions and every initial value problem $\phi_{j,\lambda} (0) =
a$, $\phi ' _{j,\lambda} (0) =b$, will have a unique solution. A
contact with the original boundary value problem is established by
imposing $\phi _{j,\lambda}(0) =0$; the condition on the
derivative is merely a normalization and for convenience we choose
$\phi_{j,\lambda} ' (0) =1$. The eigenvalues for the boundary
value problem are then discovered by imposing \beq\phi_{j,\lambda
} (L) =0,\label{imeq} \eeq considered as a function of $\lambda$.
To see a little better how this works, consider the case $R_2 =0$.
The unique solution of the initial value problem described is
\beq\phi_{2,\lambda} (\tau ) = \frac{\sin (\sqrt{\lambda} \tau )
}{\sqrt \lambda}. \nn\eeq Eigenvalues follow precisely from the
condition $$\phi_{2,\lambda} (L) =0.$$ But having the implicit
eigenvalue equation (\ref{imeq}) at our disposal, the calculation
of the determinant is basically done! Arguing as below (\ref{5})
we write \beq \zeta _{P_1} (s) - \zeta _{P_2}(s) &=& \frac 1 {2\pi
i} \int\limits_\gamma d\lambda \,\, \lambda^{-s} \frac  d {d
\lambda}\ln \frac{\phi_{1,\lambda} (L)} {\phi_{2,\lambda} (L)}
\nn\\ &=& \frac{\sin \pi s} \pi \int\limits_0^\infty dx \,\,
x^{-s} \frac d {dx} \ln \frac{\phi_{1,-x} (L)}{\phi _{2,-x} (L) }
,\nn\eeq valid about $s=0$ because the leading behavior as
$x\to\infty$ of $\phi_{j,-x} (L)$ does not depend on the potential
$R_j (\tau )$; as evidence see the analysis in Section 3. So as
around (\ref{simple})
$$\zeta _{P_1} ' (0) - \zeta _{P_2} ' (0) =- \ln \frac{\phi _{1,0}
(L)} {\phi_{2,0} (L)} $$ and we obtain the Gel'fand-Yaglom formula
\cite{gelf60-1-48}
$$\frac{\det P_1}{\det P_2} = \frac{\phi_{1,0} (L)} {\phi_{2,0}
(L)}. $$ The ratio of determinants is determined by the boundary
value of the solutions to the homogeneous initial value problem $$
\left( - \frac{d^2}{d\tau ^2} + R_j (\tau ) \right) \phi_{j,0}
(\tau ) = 0 , \quad \quad \phi_{j,0} (0) = 0 , \quad \phi_{j,0 } '
(0) = 1.$$ Even if no analytical knowledge about the boundary
value might be available, they can easily be determined
numerically.

\section{Conclusions}
The main aim of this contribution was to show that the analysis of
functional determinants for a large class of operators is
accessible to advanced undergraduate students. The only
prerequisites are elementary ordinary differential equation theory
and a basic course in complex variables. The beauty of the
approach is that it is easily adapted to different situations. We
have indicated how other boundary conditions than Dirichlet ones
can be dealt with. Indeed, general boundary conditions can be
considered along the same lines and generalizations of the
Gel'fand-Yaglom formula can be obtained. \cite{kirs04-37-4649}

We have mentioned that the presence of zero eigenvalues adds some
extra complication. The reason is that when deforming the contour
to the negative real axis, some contribution from the origin may
result. But again, a minor modification of the procedure allows
for a complete analysis. \cite{kirs03-308-502}

Even systems of differential equations can be considered with
about the same effort. \cite{kirs04-37-4649}

An example where all of the above generalizations need to be
considered is the study of transition rates between metastables
states in superconducting rings. For this case, a $2\times
2$-system with twisted boundary conditions needs to be analyzed;
see, e.g., Refs. 18 and 26\nocite{tarl94-49-494,kirs03-308-502}.

The many advantages of this approach described show that it is
optimally adapted to the evaluation of determinants. Instead of
struggling with the needed mathematical manipulations the students
should be able to easily get a grasp of this technique and to
concentrate on the underlying physics.

\section*{Acknowledgments} KK acknowledges support by the
Baylor University Summer Sabbatical Program and by the Baylor
University Research Committee.

\end{document}